\documentclass[twocolumn,amsmath,floats,showpacs,nofootinbib]{revtex4}
\usepackage[usenames]{color}
\usepackage{graphicx}
\usepackage{epstopdf}
\usepackage{textcomp}
\usepackage{hyperref}
\usepackage{multirow}
\usepackage{tikz}
\usepackage{rotating}
\hypersetup{colorlinks=true}

\begin{document}

\title{Size-effect of Kondo scattering in point contacts
(revisited)}

\author{I. K. Yanson,  V. V. Fisun, N. L. Bobrov, J. A. Mydosh*, J. M. van Ruitenbeek*}
\affiliation{B. Verkin Institute for Low Temperature Physics and Engineering, 47, Lenin Ave., 310164 Kharkov, Ukraine\\
Kamerlingh Onnes Laboratorium, Leiden University, P.O. Box 9506, 2300 RA Leiden, The Netherlands*\\
Email address: bobrov@ilt.kharkov.ua}
\published {\href{http://fntr.ilt.kharkov.ua/fnt/pdf/24/24-7/f24-0654e.pdf}{Fiz. Nizk. Temp.}, \textbf{24}, 654 (1998); [\href{http://dx.doi.org/10.1063/1.593630}{Low Temp. Phys.} \textbf{24}, 495(1998)]}
\date{\today}

\begin{abstract}The size-effect of Kondo-scattering in nanometer-sized metallic point contacts is measured with the simplified, mechanically-controlled break-junction technique for CuMn alloy of different Mn concentrations: 0.017; 0.035; and 0.18 ($\pm$0.017) at.\%. The results are compared with our previous publication on nominally 0.1 at.\% CuMn alloy \cite{1,2}. The increase of width of the Kondo resonance and enhanced ratio of Kondo-peak intensity to electron-phonon scattering intensity is observed for contacts with sizes smaller than 10 $nm$. From the comparison of electron-phonon scattering intensity for the pressure-type contacts, which correspond to the clean orifice model, we conclude that the size effect is observed in \emph{clean} contacts with the shape of a \emph{channel} (nanowire).


\end{abstract}

\maketitle
\section{INTRODUCTION}
Recently the size-effect of Kondo scattering has been observed in point contacts \cite{1,2}. This appears to be opposite to the suppression of the Kondo effect in thin films and wires \cite{3}. The effect was observed in ballistic contacts of nanometer-size and explained as being due to the strong enhancement of Kondo temperature by fluctuations of local electron density of states. These fluctuations are the result of the lateral electron resonances in the narrow part of the metallic bridge connecting bulk electrodes \cite{4}. In the present work we have found additional confirmation of the model, proposed in Ref.\cite{4}.

\begin{figure}[]
\includegraphics[width=8.5cm,angle=0]{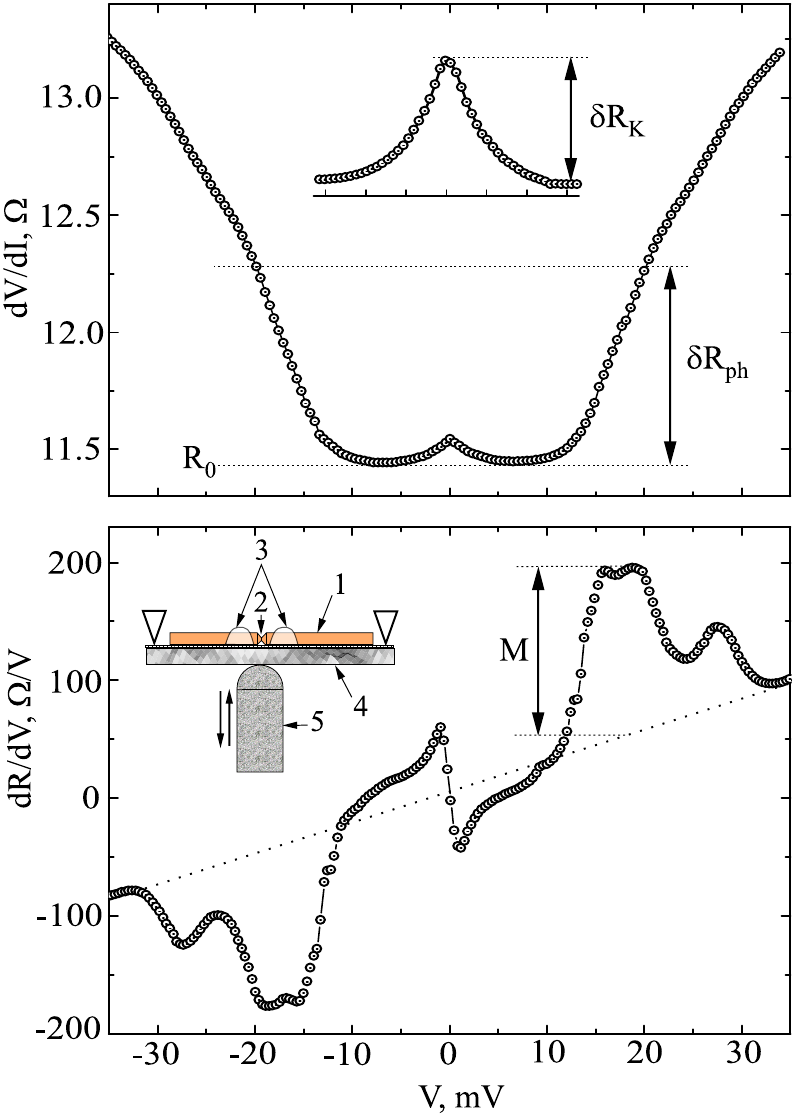}
\caption[]{Upper panel: differential resistance of MCB junction of CuMn (0.18 at.\%). In the inset the magnified Kondo peak is shown. $R_0$ is a resistance connected with contact diameter $d$ via Sharvin formula (Eq.(\ref{eq__4})). $\delta R_{ph}$ is the increase of resistance due to the phonon backscattering at $V = 20\ mV$. $\delta R_{K}$ is Kondo-peak height. Lower panel: the second derivative of current-voltage characteristic of the same contact. M is the maximum intensity of electron-phonon interaction background subtracted, according to formula (Eq. (\ref{eq__3})). In the inset: the schematic view of mechanically-controlled break junction: 1 - the CuMn alloy; 2 - notch; 3 - Staycast glue; 4 - bending beam; 5 - push-pulling rod, $T = 1.6\ K$.}
\label{Fig1}
\end{figure}

Initially, it was noticed that the estimates of the Kondo-temperature made by the quasi-classical theory \cite{1,2,5,6} is too large compared to the bulk value. Unfortunately, the experimental conditions do not satisfy the weak coupling limit in Kondo scattering which is assumed in the theory. Another problem is an unknown geometry and electron mean free path in a constriction, which are often encountered in the point-contact study and which force one to use the idealized clean-orifice model \cite{7}. Roughly speaking, due to this model any dependence of a normalized Kondo-peak intensity at zero bias $\delta R_K/R_0$ (Fig.\ref{Fig1}) on a contact diameter $d$ slower than proportional to $d$ would result in the increase of the apparent Kondo temperature according to the formulae (\ref{eq__1})

\begin{equation}
\label{eq__1}
k_BT_K=E_F \text{exp}(-2E_F/3J) ,
\end{equation}
\begin{equation}
\label{eq__2}
\frac{J}{{{E}_{F}}}=-0.044{{\left[ \frac{1}{c\sqrt{{{R}_{0}}}}\ \frac{d{{R}_{K}}}{d({{\log }_{10}}V)} \right]}^{1/3}}\ ,
\end{equation}

since $\delta (\text{log}\ eV)$ are approximately constant and $d\propto R_0^{-1/2}$. Here $E_F$ and $c$ are the Fermi energy and
impurity concentrations, respectively. There are many reasons for the exponent $n$ in $\delta R_K/R_0\propto d^n$  to be smaller than one. One of them is the finite electron mean path\footnote{In the limit of dirty contact $n = 0$ \cite{7}}, the other is changing the contact shape while studying the $d$-dependence. The theory \cite{6} is valid in the limit when the average distance between magnetic impurities $r_0$ is much smaller than the contact size $d$ which is not the case for the smallest contacts. Hence, the application of formula (\ref{eq__2}) can't be justified.

Fortunately, for $r_0 \ll d$ the ratio of size-dependent phonon and Kondo scatterings does not depend on the geometry and mean free path in the contact, and can be taken as a reliable evidence of different behavior of these two types of scatterers. Another important experimental feature is the widening of the Kondo peak while decreasing the contact size $d$ \cite{1,2} which also qualitatively points to an increase of the Kondo temperature.

In the present work we use these properties to show qualitatively that the Kondo temperature indeed greatly increases for nanometer-sized contacts in the form of a clean channel (wire). We prove that the clean channel (wire) model is essential for clear observation of the effect. In the previous publication \cite{8} we have noted this feature, but only in the present work (based on a great amount of
experimental data) we do find it to be the necessary condition for observing the size effect. The effect is shown for different known concentrations, and this enables us to correct our previous results for CuMn with nominal concentration 0.1 at.\% \cite{1,2}, which corresponds to the measured concentration of 0.028 at.\% (uncertainty $\pm$0.017 at.\%). The size-effect is maximal when the wire diameter is of the order or less than the average distance between impurity, and decreases with shortening of the electron mean free path. These conclusions correspond pretty well to the Zarand-Udvardi theory \cite{4}.

A more direct way of showing the enhanced Kondo temperature is to study nanosize contacts in magnetic fields. It was shown that Kondo resonance becomes less sensitive to the field for decreased sizes \cite{1,2,9,10}. Despite the experimental difficulties of preserving high resistance contacts during magnetic field measurements, these experiments would give not only qualitative but also the quantitative information. These remain for future studies.

\section{Experimental}
We study CuMn dilute alloys of three concentrations determined by $x$-ray analysis: 0.18; 0.035; and 0.017 at.\%. In addition, we repeat the measurements for our previous alloy which was nominally about 0.1 at.\% \cite{1,2} but appears to be 0.028 at.\% by $x$-ray. The accuracy of concentration determination is about 0.017 at.\%. Thus, the alloys with concentration of 0.017; 0.028; and 0.035 at.\% give the same results in the limits of data-point scattering.

The measurements were carried out on the break junctions shown schematically in the inset of Fig. \ref{Fig1} (lower panel). The sample (1) is a wire with diameter 0.2~$mm$ notched (2) at the center by a sharp knife, then etched and annealed at $\rm 700^\circ C$ during 2,5 h with spontaneously cooling down. The annealing seems to be important for obtaining elongated neck while breaking the notch. The sample is glued to the substrate (4) with Staycast (3). The rod (5) is pushed mechanically to bend the substrate (4). The whole system is immersed in superfluid He at a temperature of 1.6~$K$. This improves the sample cooling and greatly simplifies the measurements which enable us to collect a huge amount of experimental data.

After breaking the neck the contact is readjusted for the initial resistance of the order of $1\ \Omega$. The continuous pulling off the electrodes (1) enable us to obtain the successive series of resistances up to several hundreds Ohm until the neck is completely broken. Each series containing about 10 contacts is repeated several times. Sometimes the contact resistance inside the same series jumps to unwanted high values. Then it was readjusted back by pushing the electrodes slightly. After a number of breakings, the metal in the contact region becomes so defective that the making of an elongated clean neck appears to be difficult. These series show a suppressed Kondo size-effect.

For each contact three successive recordings are taken, each of about 5-10 minutes long. These are the $d^2V/dI^2(V)$ and $dV/dI(V)$ characteristics taken in the range of phonon energies $-$35-35~$mV$, and $dV/dI(V)$ taken at about $-$8-8~$mV$ near zero bias for Kondo-peak recorded separately (see Fig.\ref{Fig1}). Usually, the non-linearities of $V(I)$ characteristics due to phonons and Kondo-effect are the same on different recordings, showing that the contact is stable during the measurements. Sometimes, especially for low impurity concentration and small size contacts, the intensity of Kondo peak changes due to the electromigration of impurities. For these cases we use either the maximum intensity (which is observed for the previous recording), or the average of two, since we assume that the maximum current density in the center of the contact forces the impurities to move from the center to the periphery \cite{11}.

Care is taken to have the temperature and modulation smearing \footnote{The smearing of the first and second derivatives are equal
to $\delta(eV_1)=[(2.45eV_1)^2+(3.53k_BT^2)]^{1/2}$	and	$\delta(eV_2)=[(1.72eV_1)^2+(5.44k_BT^2)]^{1/2}$, respectively, according to \cite{7,24} $V_1$, $T$ are the effective values of modulation voltage and temperature, respectively.} less than the changes of the measurable quantities discussed below.
\section{Results}
Typical first and second derivatives of current-voltage characteristic are shown in Fig.\ref{Fig1}. The Kondo peak at zero bias is seen on the $dV/dI(V)$-characteristic. The phonon backscattering sharply increases the resistance at about $\pm 20\ mV$. In the ballistic limit the $dR/dV(V)$ characteristic is directly proportional to the electron-phonon interaction (EPI) function $g_{PC}(eV)$ through the relation \cite{7}

\begin{equation}
\label{eq__3}
\frac{1}{R}\frac{dR}{dV}(V)=\frac{8}{3}\ \frac{ed}{\hbar {{v}_{F}}}{{g}_{PC}}(eV)\ ,\quad (T\simeq 0).
\end{equation}

The Sharvin resistance $R_0$ which we identify with shallow minima in $dV/dI(V)$ is connected in copper with contact diameter d through the formula
\begin{equation}
\label{eq__4}
d\approx \frac{30}{\sqrt{{{R}_{0}}\left[ \Omega  \right]}}\ \left[ nm \right].
\end{equation}
This formula is valid both for the model of orifice and for long wire only in the case of specular reflections from inner boundaries. We shall use $d$ as a parameter characterizing the contact diameter although it should be remembered that it is model dependent in the general case. We recall that the true experimental parameter is the contact resistance $R_0$ which can be restored through Eq.(\ref{eq__4}).

As the quantitative estimates of phonon and Kondo scattering intensity we choose the increase of differential resistance from $R_0$ to the value at $V=\pm20\ mV$ $(\delta R_{ph})$ and the height of Kondo peak $(\delta R_{K})$ shown in Fig.\ref{Fig1}, upper panel, respectively. One more parameter proportional to the phonon scattering is the intensity of maximum M in $dR/dV(V)$ spectra which can be connected with the maximum of EPI function through Eq.(\ref{eq__3}). This quantity has an advantage that the background [approximated by a straight line (shown in the lower panel of Fig.\ref{Fig1})] can be subtracted.
\subsection{Contact-size dependence of phonon and Kondo scatterings}
\begin{figure}[]
\includegraphics[width=8.5cm,angle=0]{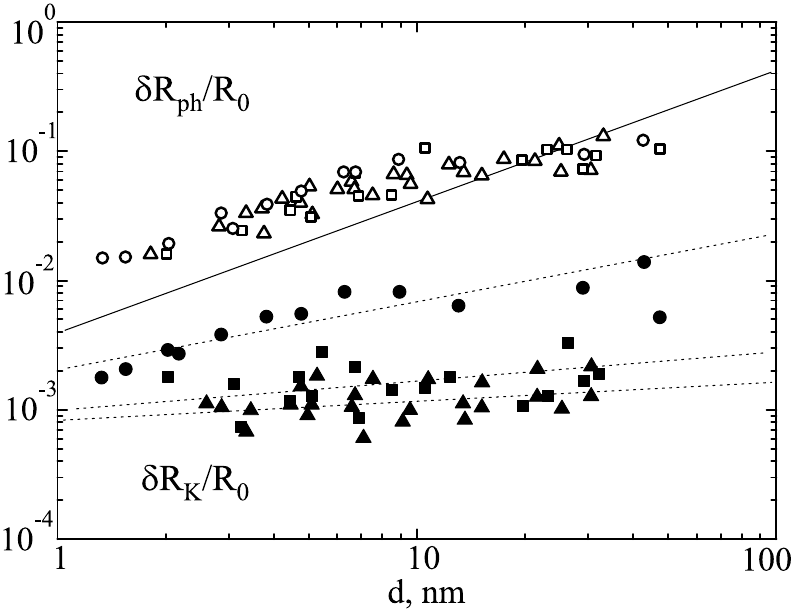}
\caption[]{Normalized increase of differential resistances due to phonon (open data points) and Kondo (solid data points) scatterings. Triangles, squares and circles stand for CuMn alloys of 0.026; 0.028; and 0.18 at.\%, respectively. The solid straight line stands for the clean-orifice phonon-scattering intensity. Dotted straight lines are the apparent linear fits for Kondo scattering of three concentrations.}
\label{Fig2}
\end{figure}

Figure \ref{Fig2} shows the cumulative results of our measurements. The phonon (open) data points follow the same trend for all concentrations, namely, at small sizes they are proportional to $d$ while for large diameters they are flattened. These new data correspond well to our previous results cited in Refs.\cite{1,2,8}. The pure copper follows the same dependence (not shown). The solid straight line denotes the $\delta R_{ph}/R_0$ vs· $d$ for a clean orifice model according to tabulated experimental $g_{PC}(eV)$ function in Ref.\cite{7}:

\begin{equation}
\label{eq__5}
{{\left( \frac{\delta {{R}_{ph}}}{{{R}_{0}}} \right)}_{V=20\;mV}}=\frac{8}{3}\frac{ed}{\hbar {{v}_{F}}}\int\limits_{0}^{20\;mV}{{{g}_{PC}}(\omega )d\omega =}
\end{equation}
\[=4.05\cdot {{10}^{-3}}d\left[ nm \right].\]

It is important that the experimental points lie above the clean orifice line for sizes smaller than 10~$nm$. For larger diameters the experimental points coincide with and lie below this line since the elastic electron mean free path becomes smaller than the contact diameter.

In general, the Kondo data points rise less steeply with increasing \emph{d} than to the phonons; this is especially clear for lower concentrations. The apparent linear fits for $\delta R_{K}/R_0$ vs. $d$ amounts\\
a)	log Y = 0.51(0.06) log X\ -2.68(0.05) for 0.18\%,\\
b)	log Y = 0.21(0.11) log X\ -2.99(0.12) for 0.028\%,\\
c)	log Y = 0.14(0.08) log X\ -3.05(0.08) for 0.026\%,\\
which are shown in Fig.\ref{Fig2} with the straight dotted lines. Here X = $d\ [nm]$, Y = $\delta R_{K}/R_0$ and in brackets we denote the standard deviation. We join the experimental data for concentrations of 0.035 and 0.017 at.\% in one set with an averages concentration 0.026 at.\%, since they are indistinguishable in view of large uncertainty (0.017 at.\%).

The scattering of data points is rather high which is due to the dispersion between different series. The scattering inside each series is much less. Thus, we can conclude that the property of the metal differs from sample to sample, and from the series to series for the same alloy.

The alloy with nominal concentrations of 0.1 at. \% which we have studied in previous works \cite{1,2,8} yields quite a similar exponent b) (in the limits of errors) dependence \footnote{Equation cited in Ref.\cite{8} reads:\\$\rm log Y = 0.16(0.05)logX-2.62(0.05).$} in the present study despite the quite different experimental conditions (T=1.6~$K$ instead of 0.5~$K$, an environment of liquid helium instead of high vacuum).

\begin{figure}[]
\includegraphics[width=8.5cm,angle=0]{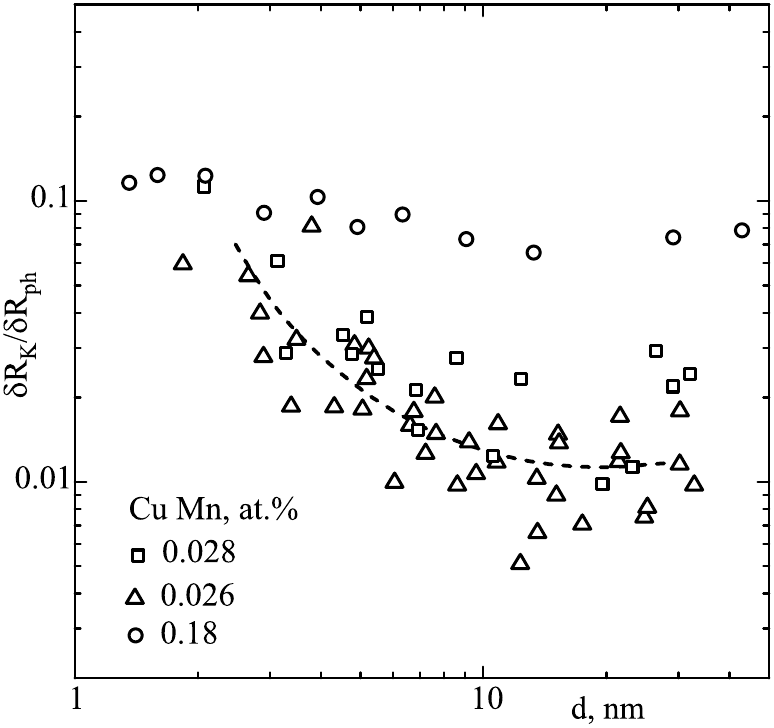}
\caption[]{The ratio of resistance increase due to Kondo and phonon scattering as a function of diameter. Squares, triangles, and circles stand for CuMn alloys of 0.028; 0.026; and 0.18 at.\%, respectively. The dashed line serves as the guide to the eye for 0.026 at.\%.}
\label{Fig3}
\end{figure}

For pure copper the zero-bias Kondo peak $\delta R_{K}/R_0$ is of the order of $1e^{-4}$ which corresponds to the purity of our Cu metal. Previous studies in the analogous break-junction devices \cite{12,13} also have not revealed the zero-bias Kondo peaks in pure copper. The ratio between Kondo and phonon intensities are plotted in Fig.\ref{Fig3}. As we have noted above this ratio does not depend upon the constriction geometry and elastic mean free path. It is seen that $\delta R_{K}/\delta R_{ph}$ increases by almost an order of magnitude for diameters smaller than 10~$nm$. For higher concentration (0.18 at.\%) this dependence is masked. Here we want to notice that, while 0.017; 0.028; and 0.035 at.\% correspond to different series and samples (and this is a cause of big scattering of data points) the results for 0.18 at.\%, represent only a single series and are more continuous.
\subsection{Energy dependence of Kondo scattering}
\begin{figure}[]
\includegraphics[width=8.5cm,angle=0]{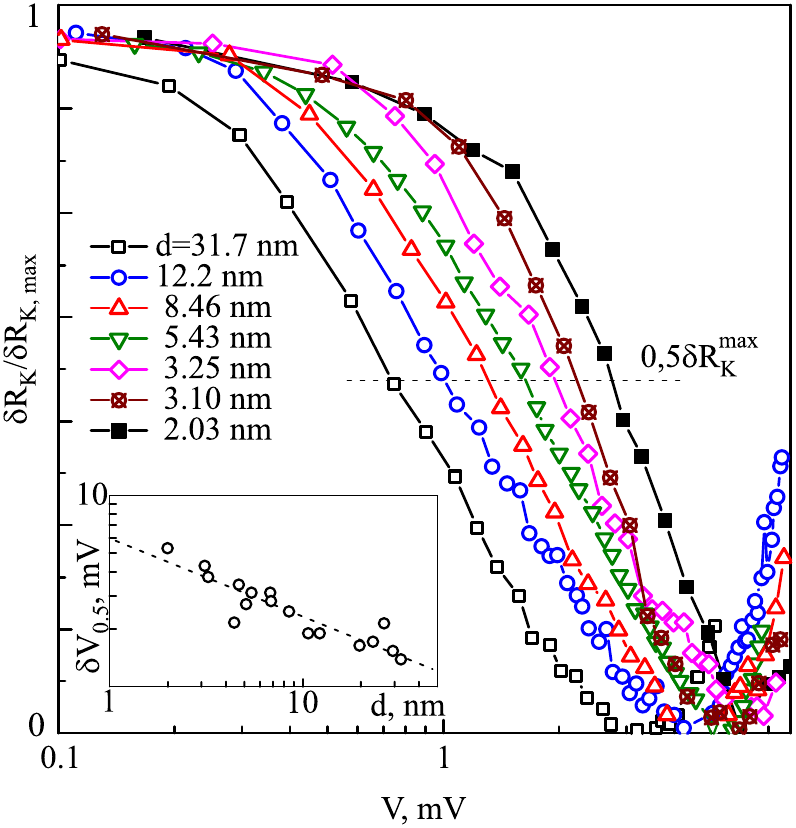}
\caption[]{The normalized Kondo maximum for CuMn (0.028 at.\%) for different contact diameters. In the inset: the width of the maximum at the half height (see the horizontal dotted line in the main panel) is shown as a function of diameter.}
\label{Fig4}
\end{figure}

\begin{figure}[]
\includegraphics[width=8.5cm,angle=0]{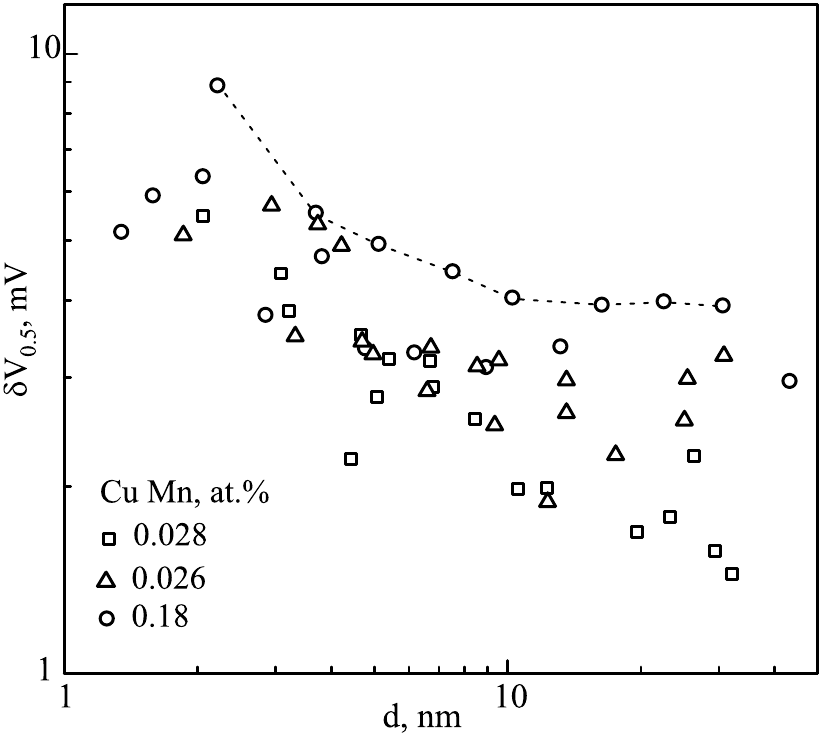}
\caption[]{The width of the Kondo maximum as a function of contact diameter. Squares, triangles, and circles stand for CuMn alloys of 0.028; 0.026; and 0.18 at.\%, respectively. The dashed line connects data-points of single series.}
\label{Fig5}
\end{figure}

The energy dependence for the normalized Kondo resistance is shown as a function of contact diameter for 0.028 at.\% alloy in Fig.\ref{Fig4}. For other concentrations these dependences look similar. The new measurements are practically the same as already reported (see Fig.\ref{Fig2}(b) in Ref.\cite{9} and Fig.\ref{Fig5}(b) in Ref. \cite{10}), but are more detailed since the simplified experimental technique enables us to collect more data. The logarithmic voltage dependences of $\delta R_{K}/\delta R_{K,\ max}$ are quite evident for all curves. In the inset to Fig.\ref{Fig4} the voltage width at the half-maximum height vs. diameter is plotted as shown with the dotted horizontal line in the main panel. More data points are plotted in the inset since not the all curves are shown in the main panel. The observed scattering is due to the data points from different series.

The widths of the Kondo-resistance-peaks are shown for different concentrations as a function of contact diameter in Fig.\ref{Fig5}. In spite of the large data scattering, the increase of $\delta V_{0.5}(d)$ for diameters
smaller than 10~$nm$ is clearly seen. This increase can be observed for each concentration, including 0.18 at.\%. It proves that even for this alloy we did observe increased energy-scale for Kondo-effect, although it may not be so evident from Fig.\ref{Fig3}. It is
important to note that the increase of Kondo-peak width was not observed in Cu(Fe)-contacts \cite{9,10} with comparable and higher resistances. The difference between Mn and Fe impurities proves that the widening of Kondo-peak is not due for extrinsic effects (like quantum diffraction of electron wave functions at small constrictions) but is connected with the Kondo-mechanism itself. Note also, that at the \emph{large} diameters the width of the Kondo-peak increases at the increasing Mn concentrations in the row 0.028; 0.035, and 0.18 at.\%, as expected due to the spin-glass effects. Indeed, for low-resistance contacts with $c = 0.0018$ we observe the splitting of Kondo-peak due to the internal field (not shown here).

\section{Discussion}
The PC EPI spectral functions are summarized for pressure-type point contacts in Refs.\cite{7,14}. These are made either by pressing together the sharp needle to the flat surface (other version: by pressing together the sharp edges of metallic electrodes) or by electrical microwelding by a current pulse. In all these cases the probability of formation of the metallic contact with a length much smaller than its width (diameter) is high enough. These contacts are satisfactorily modelled as an orifice in an infinitely thin partition. Indeed, the experimentally observed maximum intensity of PC EPI spectra are saturated at a constant value. For copper this value is about $g_{PC}^{max}= 0.24$ (linear background being subtracted). Theoretical calculation for the
simplest metal - sodium - gives a value coinciding with the experiment, that can be taken as a quantitative proof of observing the clean orifice \cite{14}. With invention of nanofabricated thin film junctions \cite{15}, STM \cite{16}, and mechanically- controllable break (MCB) junctions \cite{17} a new possibility appears. The shape of a contact can be fabricated as a channel (wire) whose length is equal or greater than a diameter. According to Ref.\cite{18}, the shape of a contact for STM (and evidently, for MCB \cite{19}) depends on the fabrication procedure, and can be made similar to both orifice and channel for different prehistories. It was shown in Ref.\cite{18} for gold STM junctions that the gentle touch of a needle to the flat surface leads to observing the contact whose shape was close to the orifice, while the more deep intrusion results in producing a long wire. Something like this one could expect also for the MCB junctions.

\begin{figure}[]
\includegraphics[width=8.5cm,angle=0]{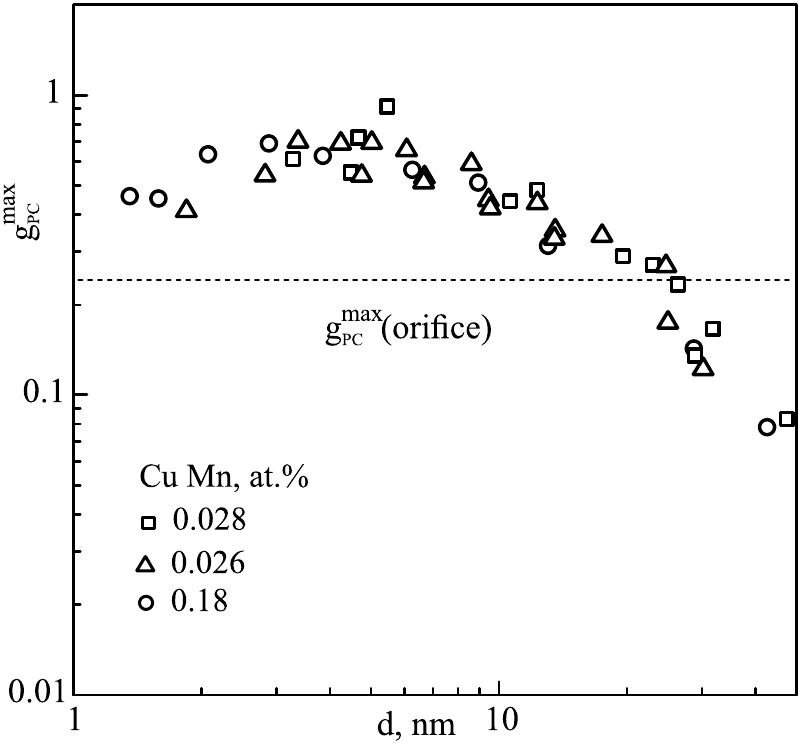}
\caption[]{Maximum of the electron-phonon interaction spectral function as a function of contact diameter. Squares, triangles, and circles stand for CuMn alloys of 0.028; 0.026; and 0.18 at.\%, respectively. Horizontal dotted line shows the maximum $g_{PC}$ of clean orifice value.}
\label{Fig6}
\end{figure}

We show the $g_{PC}^{max}$ for our contacts of different concentration in Fig.\ref{Fig6}. It is seen that the maximum intensities exceed the orifice value by 2-3 times. This was never observed for pressure-type contacts. If we assume a linear interpolation between theoretical limits for clean orifice and channel \cite{7}

\begin{equation}
\label{eq__6}
\frac{1}{R}\frac{dR}{dV}(V)=\frac{8}{3}\frac{e\left( d+\frac{3\pi }{4}L \right)}{\hbar {{v}_{F}}}{{g}_{PC}}(eV),\ (T=0),
\end{equation}
then this means that the length of the "wire" (or channel) is approximately equal to its diameter.

Correspondingly, the experimental data points for $\delta R_{ph}/R_0$ in Fig\ref{Fig2} lie about twice as high than
the straight solid line which stands for the clean orifice model according to the Eq.(\ref{eq__5}). Thus, we can state that our contacts with diameter smaller than $\simeq 10\ nm$ are similar to nanosized wire (channel). More about the length, shape, and purity of this wire can be said from the $d$-dependence of phonon intensity but this lies beyond the scope of this work.

\begin{figure}[]
\includegraphics[width=8.5cm,angle=0]{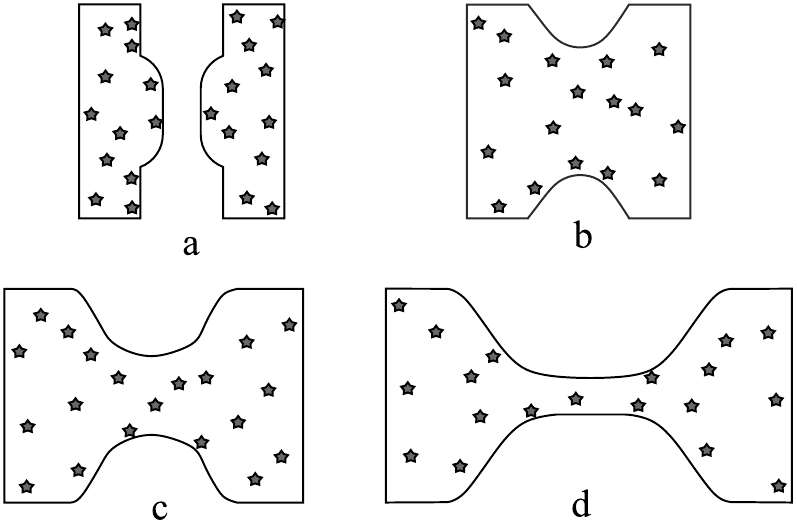}
\caption[]{Schematic view of MCB junction: before touching (a); first touch starting from the low resistance contact (b); successively forming the neck while pulling the electrodes off ( c and d). Small stars denote the magnetic impurities.}
\label{Fig7}
\end{figure}
Figure \ref{Fig7} shows schematically what may happen. After the first breaking there are presumably hillocks seeing each other (a) and separating by the least distance. First, the pressure type contact gives the shape more like an orifice (b). These correspond to contact diameter greaters than 20~$nm$. The more pressure - the more defects are introduced, and this leads to a decrease in the phonon intensity below the clean orifice value (Fig.\ref{Fig6}) for the largest diameters. Pulling off the electrodes results to a shape similar to a wire (channel), but the greater the separation distance the more defects are introduced to the thinner wire, and this probably leads to a decrease in the intensity at the least contact diameters. Other causes may be quantum diffraction effects for electron and phonon, since their wavelengths become comparable to the wire diameter \cite{20}.

Recently, Kolesnichenko et al. \cite{21} have predicted a classical mesoscopic size effect for impurities whose average interimpurity distance is comparable or greater than the contact diameter. This effect is a version of the classical correlation phenomena in point contacts considered by Gal'perin
and Kozub in Ref.\cite{22}. Due to the scattering probability which is proportional to the solid angle by which an orifice is seen from the impurity location, the maximum scattering probability has the impurity located at the interface of the orifice model. It is easy to see that the number of these impurities increases if the shape of the contact deviates from an orifice to a wire (Fig.\ref{Fig7}). Qualitatively, this effect can explain nonlinear in $d$ dependence of Kondo scattering (with increase at small size contacts), but it hardly can be responsible neither for the increase of width of the Kondo peak (Fig.\ref{Fig5}), nor for the loss of sensitivity to external magnetic field for ultra small contacts.
\section{Conclusion}
The results of the present work proves the correspondence of Zarand-Udvardi model \cite{4} to the experiments. This model ascribes the enhancement of the Kondo temperature of magnetic impurities in metallic point contacts due to fluctuations of the local electron density of states. There are no fluctuations for a pure orifice since lateral quantization of electron wave function is smeared out \cite{23}. On the other hand, for dirty channels (wires) quantization is destroyed by scatterings, especially if they involve spin flip. Note, that for large biases (of the order of phonon characteristic frequencies) Kondo scatterings with spin flip are negligible. Thus, we conclude that the size effect can be observed in clean enough nanosized wires, in accordance with Ref.\cite{4}. The necessary conditions for observing the Kondo size effect in point contacts are that the phonon intensity should be noticeably greater than that of a clean orifice model \cite{7}.

We are indebted to A.~I.~Yanson who settled the automatic data recording in our lab, which enabled us to carry out this work. Our thanks are to V.V.~Demirski for making the quantitative analysis of alloys. I.~K.~Y. and N.~L.~B. appreciate the financial support of Soros Foundation. I.~K.~Y. is grateful to Alexander von Humboldt Foundation and to the Physikalishes Institute of Karlsruhe University for hospitality in the frame of Humboldt-Forschung- spreistrager Program.
The authors are delighted to dedicate this paper to Academician I.M.~Dmitrenko on the occasion of his 70th birthday.

\end{document}